\documentclass[epj,twocolumn]{svjour}
\usepackage{amsfonts}
\usepackage{amsmath}
\usepackage{amssymb}
\usepackage{graphicx}
\usepackage{xspace}
\usepackage{color}
\usepackage{float}

\newcommand{\sinc}{{\rm sinc}} 
\begin{document}


\title{Graphene with time-dependent spin-orbit coupling: 
Truncated Magnus expansion approach}
\author{A. L\'{o}pez\inst{1}\and A. Scholz\inst{1}\and Z. Z. Sun\inst{2} \and J. Schliemann\inst{1}}
\institute{Institute for Theoretical Physics, University of
Regensburg, D-93040
Regensburg, Germany\and
School of Physical Science and Technology, Soochow University, 
Suzhou, Jiangsu 215006, China
}
\mail{alexander.lopez@physik.uni-regensburg.de}
\titlerunning{Graphene with time-dependent spin-orbit coupling}
\abstract{We analyze the role of ac-driven Rashba spin-orbit coupling in 
monolayer graphene including a spin-dependent mass term. 
Using the Magnus expansion as a semi-analytical
approximation scheme a full account of the quasienergie spectrum of spin states is given. 
We discuss the subtleties arising in correctly applying the 
Magnus expansion technique in order to determine the quasienergy spectrum. 
Comparison to the exact numerical solution gives appropriate 
boundaries to the validity of the Magnus expansion solution.
}
\maketitle
\section{Introduction}

Typically, the dynamics of time-periodically driven systems is solved by 
means of Floquet's theorem.\cite{milena,chu} A standard approach 
consists in performing a Fourier mode expansion of the eigenstates and then deal 
with the corresponding infinite dimensional eigenvalue 
problem by means of either numerical or approximate strategies to determine 
the so-called quasienergy spectrum. One alternative route which avoids 
the infinite dimensional eigenvalue formulation 
was put forward by Magnus \cite{magnus1,magnus2} who proposed 
an exponential solution 
for the evolution operator given in terms of 
nested commutators of the time-dependent Hamiltonian. In a previous work 
 \cite{lopez} some of us have used the Magnus expansion (ME) in order to 
analyze the role of an ac-driven Rashba spin-orbit coupling (RSOC) on the 
spin dynamics of charge carriers in single 
layer graphene. However, as it was
pointed out  recently  by Zhou and Wu \cite{arxiv}, the reported 
results 
where beyond the convergence region of the ME solution. 
This in turn stems from the fact that within the Schr\"{o}dinger picture 
representation, the convergence of the ME solution is 
momentum-dependent and, as a result, only valid for low momenta of the charge 
carriers. Indeed, in reference \cite{arxiv} it was shown that in 
the Schr\"{o}dinger picture the ME solution is only valid 
within a small neighborhood of the Dirac point.

In the present paper we perform a thorough analysis of the subtleties 
arising when implementing the ME as a semi-analytical
approach to the above quantum dynamics. As we shall see, 
switching from the Sch\"odinger to the interaction picture significantly 
improves the convergence behavior.\cite{feldman,fernandez,fme} 
This improvement, however, comes at a price: The Hamiltonian in the
interaction picture is no longer time-periodic. As a consequence,
the resulting time evolution operator in the Schr\"odinger representation 
arising from a
{\em truncated} Magnus expansion does in general not  fulfill the so-called
stroboscopic property required from Floquet formalism.\cite{milena,chu} 
However, as we shall see in the following, despite this shortcoming
the results of the truncated analytical Magnus expansion performed in the 
interaction picture agree well with the exact numerics within the convergence range.

The paper is organized as follows. In section {\rm II} we introduce the model 
Hamiltonian for periodically driven RSOC and present its exact analytical 
solution right at the Dirac point. The main results of the 
Floquet-Magnus approach for the semi-analytical 
solution of the evolution operator  are presented in 
section {\rm III}. Here, we also explore the ME for different bounds on the 
convergence domain reported in 
the literature. Next, in section {\rm IV} we apply the ME approach to the 
driven RSOC problem at finite momentum, and compare 
with the 
exact numerical solution of the quasienergy spectrum. In section {\rm V} 
we present a discussion of our main results. Finally, in 
section {\rm VI} we give some concluding remarks and outlook.

\section{Model}
\label{model}

We consider a graphene monolayer being subject to 
a periodic time-dependent spin-orbit interation of the (extrinsic) Rashba type. 
The sample is located on a certain substrate which induces a (mass) gap in the energy 
spectrum due to the intrinsic spin-orbit coupling.
Thus, in Dirac cone approximation, the system is described by the 
 $4\times4$ Hamiltonian\cite{kane}
\begin{equation}\label{dirac1}
\mathcal{H}(\vec{p},t)
=v_F\vec{\sigma}\cdot\vec{p}+\Delta\sigma_zs_z
+\lambda(t)\hat{z}\cdot[\vec{\sigma}\times\vec{s}],
\end{equation}
where we have concentrated on the ${\bf K}$ corner point of
the Brilloiun zone being the reference of the momentum $\vec p$. The results 
for the ${\bf K'}$ point are obtained by setting $p_x\rightarrow-p_x$.
Here $v_{F}\sim 10^6 {\rm m/s}$ is the Fermi velocity and the 
Pauli matrix vectors $\vec{\sigma}$ and $\vec s$ describe the 
sublattice degree
of freedom and the electron spin, respectively. On the other hand,
$\Delta$ parametrizes the intrinsic spin-orbit coupling
and $\lambda(t)$ describes the time-dependent RSOC which can,
in principle, be induced by capacitor plates parallel to the setup
and coupled to an LC circuit. In what follows we  assume 
a time dependence of the form $\lambda(t)=\lambda_R\cos\Omega t$, 
with $\Omega=2\pi/T$ being the frequency and $T$ the period of the driving 
field. In the following we set $\hbar=1$. 

Upon applying the transformation
\begin{small}
\begin{equation}\label{unitary} 
U(\vec{p})=
\frac{1}{\sqrt{2}}\left(
\begin{array}{cccc}
e^{-i\phi}\cos\frac{\gamma}{2}&-e^{-i\phi}\sin\frac{\gamma}{2}
& e^{-i\phi}\cos\frac{\gamma}{2}&-e^{-i\phi}\sin\frac{\gamma}{2} \\
\sin\frac{\gamma}{2}&\cos\frac{\gamma}{2}&\sin\frac{\gamma}{2}
&\cos\frac{\gamma}{2} \\
i\sin\frac{\gamma}{2}&i\cos\frac{\gamma}{2}&-i\sin\frac{\gamma}{2}
&-i\cos\frac{\gamma}{2} \\
ie^{i\phi}\cos\frac{\gamma}{2}&-ie^{i\phi}\sin\frac{\gamma}{2}
&-i e^{i\phi}\cos\frac{\gamma}{2}&ie^{i\phi}\sin\frac{\gamma}{2} 
\end{array}
\right),
\end{equation}
\end{small}with $\tan\phi=p_y/p_x$ and $\tan\gamma=v_F p/\Delta$,
the time-dependent Hamiltonian (\ref{dirac1})
becomes block-diagonal
\begin{equation}\label{interaction}
\mathcal{H}(p,t)=\left(
\begin{array}{cc}
 H_-(p,t)&0 \\
0 &H_+(p,t)
\end{array}
\right),
\end{equation}
where
\begin{eqnarray}\label{h1}
\nonumber H_\pm(p,t)&=& \pm\lambda(t)\sigma_0+\frac{\Omega_0}{2}\sigma_z\\
&&\mp\lambda(t)[\cos\gamma\sigma_z-\sin\gamma\sigma_x],
\end{eqnarray} 
with $\Omega_0=2\sqrt{(v_Fp)^2+\Delta^2}$ and $\sigma_0$ being the
$2\times2$ unit matrix.
The transformation (\ref{unitary}) can be constructed by first diagonalizing
(\ref{dirac1}) at $\lambda=0$ and expressing the full Hamiltonian in that
eigenbasis. A further elementary transformation (diagonalizing the
diagonal $2\times2$ blocks of the resulting Hamiltonian matrix) then leads
to the basis encoded in (\ref{unitary}). This result simplifies
(but is equivalent to) the approach given previously in Ref.~\cite{lopez}
in that it avoids explicit reference to an initial time.

Since both subblocks just differ in the sign of $\lambda$ we
focus on $H_+(p,t)\equiv H_S(p,t)$ and treat this time-periodic
Hamiltonian via the Floquet theorem.\cite{milena,chu} 
It states that the general solution of the dynamics can be written as 
\begin{equation}\label{floquet}
U(t)=P(t)e^{-i H_F t},
\end{equation}
with $P(t)$  being a periodic and $H_F$ a constant matrix. 
The time evolution operator in (\ref{floquet}) satisfies the stroboscopic 
property
\begin{eqnarray}\label{stroboscopic}
U(nT)&=&P(nT)e^{-i n TH_F}\nonumber\\
 &=&[U(T)]^n.
\end{eqnarray}
The eigenvalues of $H_F$ determine the quasienergy spectrum of the 
periodically 
driven problem.

First let us note that an exact solution for the spin dynamics generated by 
$H_S(p,t)$ can be found at $p=0$ where
the time evolution operator explicitly reads
\begin{equation}\label{p0}
U(t)=\mathcal{P}(t)e^{-i\Delta t\sigma_z},
\end{equation}
with $\mathcal{P}(t)=e^{-i(\sigma_0-\sigma_z)f(t)}$ and we have defined
\begin{equation}\label{f}
f(t)=\int^t_0 \lambda(t')dt'. 
\end{equation}  
Here, the quasienergy spectrum (modulo $\Omega$) is given by
\begin{equation}
 \varepsilon_\pm(p=0)= \pm\Delta.
\end{equation}
At finite momentum the evolution operator admits,  to the best of our knowledge, no analytical 
solution. A standard approach is to expand the solutions in an
appropriately truncated basis of Fourier modes and to numerically
diagonalize the resulting hermitian matrix.
A semi-analytical alternative to this procedure will be discussed now.

\section{Magnus expansion: General}

Contrary to the numerical Fourier-Floquet solution, the Magnus 
expansion (ME) 
approach \cite{magnus1,magnus2} avoids the diagonalization of a truncated
eigenvalue problem, and similarly to the Fourier mode expansion approach it also has the physical 
virtue of preserving unitarity of the time evolution to any order in the 
expansion. This has to be contrasted to the approximate solution obtained 
through the Dyson series, where truncation at any given order 
leads to a non-unitary evolution. 
Although the Dyson series expansion always converges for bounded dynamical 
generators,\cite{dyson1} this is in general not true for the Magnus series. 
Thus, our next 
task is to summarize the most relevant subtleties that arise concerning the 
convergence of the Magnus series. These convergence restrictions will be 
important for the application of the 
ME strategy to the RSOC driven problem that is presented in the next section. 
For a more detailed and general discussion we refer the reader to
reference \cite{magnus2}.  

Following Magnus \cite{magnus1}, the time-evolution operator generated by the
Schr\"odinger equation
\begin{equation}\label{flo2}
i\partial_\tau U(\tau)=H(\tau)U(\tau).
\end{equation}
(with $\tau=\Omega t$ and $H(\tau)=H_S(\tau)/\Omega$) 
can be formulated as 
\begin{equation}\label{m1}
 U(\tau)=e^{-iM(\tau)},
\end{equation}
where the exponent $M(\tau)$ is expressed as an infinite series 
\begin{equation}\label{expansion}
M(\tau)=\sum_{j=1}^\infty M_j(\tau)\,.
\end{equation}
The term $M_j(\tau)$ in this expansion  
is given in terms of integrals over
sums of nested commutators with the first contributions reading as
\begin{eqnarray}\label{m01}
\nonumber M_1(\tau)&=&\int_0^\tau H(\tau_1)d\tau_1\\\nonumber
M_2(\tau)&=&\frac{1}{2}\int_0^\tau d\tau_1\int^{\tau_1}_0[H(\tau_1),H(\tau_2)] d\tau_2\\\nonumber
M_3(\tau)&=&\frac{1}{6}\int_0^\tau d\tau_1\int_0^{\tau_1}d\tau_2\int^{\tau_2}_0\big([H(\tau_1),[H(\tau_2),H(\tau_3)]]\\\nonumber
&&+[H(\tau_3),[H(\tau_2),H(\tau_1)]]\big) d\tau_3\\
&&\vdots\nonumber
\end{eqnarray}

A key question regarding this approach is of course the convergence
of the series (\ref{expansion}).
In his original work, Magnus gave an {\em a posteriori} 
convergence criterion in terms of the eigenvalues of the resulting operator
$M$  given in Eq.~(\ref{m1}). However, for practical use
{\em a priori} criteria dealing directly with the Hamiltonian
appear also desirable. In general, the ME should not be expected to 
converge unless $H$ is small in a suitable sense compared to the other typical energy scales in the problem. 
Specifically, bounds of the form 
\begin{equation}\label{bound}
\int^\tau_0 d\tau_1\|H(\tau_1)\|_2< r
\end{equation}
have been investigated as convergence conditions, where
$\|H\|_2$ is the euclidean norm of $H$ defined
as the squared root of the largest eigenvalue of the positive semi-definite 
operator $H^\dagger H$. The convergence radius $r$ in the 
above inequality restricts the times $\tau'\in[0,\tau]$ for which
the ME is applicable. 
The task of estimating $r$ has a long history in the literature
\cite{agrachev,vela,pechukas,blanes}. Finally, it was established that
 a sufficient criterion for the convergence of the Magnus expansion 
is given by\cite{moan,casas}
\begin{equation}\label{largest}
\int^\tau_0 d\tau_1\|H(\tau_1)\|_2< \pi.
\end{equation}
To get an insight into the meaning of this convergence boundary we summarize the analysis given in {\bf lemma 3} of reference \cite{magnus2} where it is established that the 
critical value $r_c=\pi$ arises from the evaluation of the poles of the derivative of the inverse exponential function, which is the Magnus expansion. These poles are given as 
$2mi$ with $m=\pm1,\pm2,\dots$, from which one gets the lowest value $r_c=\pi$.
Moreover, the above value $r_c=\pi$ was shown to be sharp in the sense that
it cannot be enlarged without further assumptions on the 
Hamiltonian.\cite{casas}

In the following section we apply the convergence restriction (\ref{largest}) 
in order to 
determine the most suitable parameter values that allow a semi-analytical 
description of the quasienergy spectrum for the RSOC driven setup.
\begin{figure}[ht]
\begin{center}
\includegraphics[height=6.5cm]{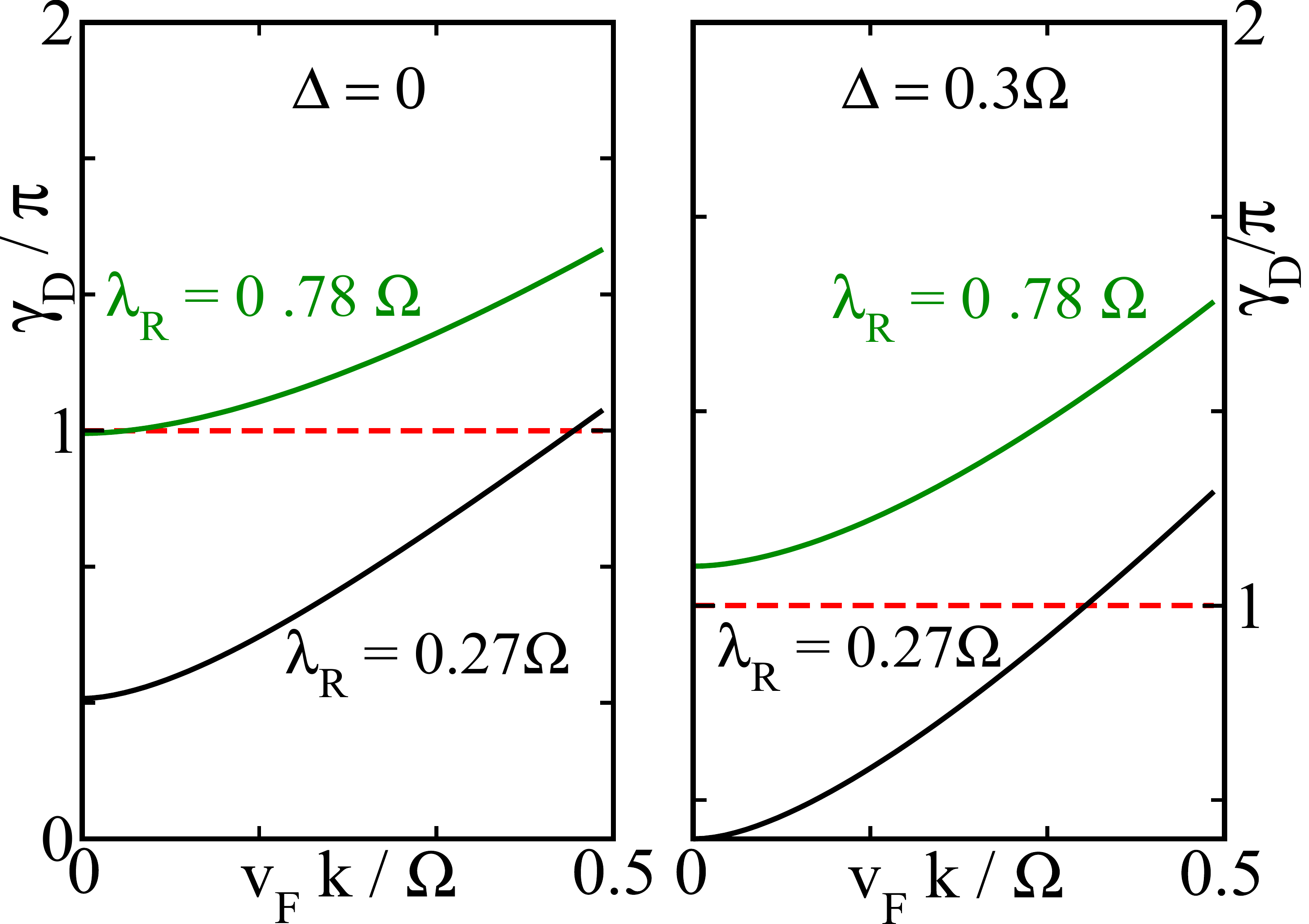}
\caption{\label{fig:figure0}(Color online) Normalized one-period 
dynamical phase  that is related to the convergence of the ME in the Schr\"{o}dinger representation 
defined as the l.h.s of inequality (\ref{convergence}) for $\tau=2\pi/ \Omega$.
The data is shown as a function of wave vector for different Rashba
coupling strength $\lambda_R$, with $\Delta=0$ (left) and
$\Delta= 0.3\Omega$ (right). The dotted red line is a guide to the eye 
corresponding to $\gamma=\pi$ in
Eq.~(\ref{convergence}). Although the convergence domain grows with
decreasing coupling parameters $\lambda_R$ and $\Delta$, the ME is, in
the Schr\"odinger representation, typically applicable only in a rather small  
neighborhood of  the  Dirac point, and for large couplings the convergence
domain indeed shrinks to zero.}
\end{center}
\end{figure}

\section{Magnus expansion: Application}

Let us first explore the criterion (\ref{largest}) in the Schr\"odinger
picture of quantum dynamics used so far.
Defining
$\kappa=v_Fp/\Omega$, 
$\Lambda=\lambda_R/\Omega$, and $\delta=\Delta/\Omega$ the
inequality (\ref{largest})
\begin{equation}\label{convergence}
\int^{\tau}_0\sqrt{\kappa^2+[\Lambda\cos\tau_1-\delta]^2}d\tau_1< \pi
\end{equation}
limits the validity of the ME solution to rather small momenta (see below), as
pointed out in Ref.~\cite{arxiv} on the basis of explicit numerics.
Another way to explore this issue is to consider the 
instantaneous eigenvalues of the time-dependent Hamiltonian,
\begin{equation}\label{dynphase}
E_\pm(\tau)= \pm\sqrt{\kappa^2+[\Lambda(\tau)-\delta]^2}, 
\end{equation}
leading to an accumulated dynamical phase
\begin{equation}
\gamma^\pm_D=\pm\int^{2\pi}_0 d\tau \sqrt{\kappa^2+[\Lambda(\tau)-\delta]^2}.
\end{equation}
in a full period (with a vanishing Berry phase).
The modulus of this expression is precisely the l.h.s of the convergence
criterion (\ref{largest}) and is plotted 
in figure \ref{fig:figure0}. As seen there, the applicability of the ME
in the Schr\"odinger picture is limited to rather small momenta.
This finding is indeed similar to earlier observations 
by Salzman.\cite{salzman2} 

A qualitative improvement is achived by changing to the 
interaction picture. Defining the ``unperturbed'' Hamiltonian $H_0$ to
be that diagonal part of $H_+(t)$ given in Eq.~(\ref{h1})
that has only a trivial time dependent term proportional to the
unit matrix, the effective ``perturbation'' $V=H-H_0$ reads in the interaction picture
\begin{equation}\label{interactionbeta}
V_I(\kappa,\tau)=\Lambda\cos{(\tau)}\left(
\begin{array}{cc}
 -\cos\gamma&\sin\gamma e^{i \omega_0 \tau} \\
\sin\gamma e^{-i\omega_0 \tau}&\cos\gamma
\end{array}
\right),
\end{equation}
with $\omega_0=2\sqrt{\kappa^2+\delta^2}$ and  $\tan\gamma=\kappa/\delta$. 
In this representation, the convergence radius of 
the Magnus solution becomes momentum-independent and restricts the 
effective coupling constant $\Lambda$ by the inequality 
\begin{equation}\label{limit}
\Lambda\int^{\tau}_0|\cos{\tau_1}|d\tau_1< \pi.
\end{equation}
Evaluation after one period gives the numerical restriction 
$\Lambda_1< \pi/4\approx 0.78$, whereas the extension to larger time domains 
reduces the coupling constant as $\Lambda_n<\Lambda_1/n$. 
Hence, it is important to notice  that in order to correctly apply the ME 
the restriction (\ref{limit}) must always be fulfilled.
Although switching to the interaction representation improves the 
convergence of the ME, the dynamical equations
\begin{equation}
i\partial_\tau U_I(\tau)=V_I(\tau)U_I(\tau)
\end{equation}
are not longer generated by a periodic operator since 
$V_I(\tau+2n\pi)\neq V_I(\tau)$. Thus, Floquet's theorem does not apply and no 
stroboscopic evolution will follow for $U_I(\tau)$. On the other hand,
the full time evolution operator in the Schr\"odinger picture
\begin{equation}\label{evolh}
 U(\tau)=e^{if(\tau)}e^{-i\omega_0\tau\sigma_z/2}U_I(\tau)
\end{equation}
 with $U_I(\tau)$ evaluated by the full (i.e., untruncated) ME of course obeys 
the stroboscopic condition (\ref{stroboscopic})
provided the ME converges. In practical
calculations, however, it is usually (as well as in the present case here) 
not possible to evaluate all terms of the ME, and one truncates this
series at rather low order.

In order to compute the quasienergy spectrum from the time evolution
operator $U(\tau)$ without making use of the stroboscopic relation
(\ref{stroboscopic}), one needs to calculate the logarithm of
$U(nT)$ for some integer $n$. The difficulty here is that the complex logarithm
is not a single-valued function, and for different $n$ the results
may lie on different sheets. Unfortunately, it is impossible to properly
keep track of these phases without prior knowledge of the
quasienergies which are in fact supposed to be the results of this
computational step. Such difficulties do not arise for $n=\pm 1$ since
the division by $n$ has only a trivial effect here, and it is easy to see
that a move to another sheet of the complex logarithm is just a shift 
to another periodicity interval of quasienergies.
Regarding general $n$, there are only very special cases where $U(nT)$
can be calculated is such an explicit fashion that one can
immediately read off the quasienergies. As an example, consider
$\gamma=\pi/2$ ($\Leftrightarrow\delta=\Delta=0$) at resonance
$\kappa=\kappa_{res}=\pi$ ($\Leftrightarrow k=\Omega/2v_F$). Here the
computational complexity is significantly reduced, and one
finds in first-order ME
\begin{equation}
U(\kappa_{res},nT)
=\exp\Big\{-i\Big(\frac{\lambda_R}{2}\sigma_x+\frac{\Omega}{2}\Big)n T\Big\}
\end{equation}
with the quasienergies $\varepsilon_\pm=\Omega/2\pm\lambda_R/2$. This
result obtained for general $n$ relies on the possibility to write
$U(nT)$ explicitly as a single exponential whose argument is linear in $n$.
Unfortunately, such a situation arises only in particular cases. Another
example is of course the case $k=0$ already discussed in section
\ref{model}.

Following the above discussion, 
we will therefore concentrate on the case $n=1$ when
evaluating the time evolution. As we shall see, this procedure leads to
accurate approximations to the quasienergy spectrum.
Since the interaction $V_I(\tau)$ is a $2\times 2$ matrix we can write $M(\tau)=\vec{m}(\tau)\cdot\vec{\sigma}$, and 
the Magnus series amounts to write the vector series 
\begin{equation}
\vec{m}(\tau)=\vec{m}_{1}(\tau)+\vec{m}_{2}(\tau)+\cdots, 
\end{equation}
where the $jth$ vector contribution is of order $\Lambda^j$ and is obtained by integration of nested commutators of the interaction term. 
The evolution operator in the interaction picture can be written as
\begin{equation}\label{interad}
U_I(\tau)=\cos{|\vec{m}(\tau)|}-i\sinc{|\vec{m}(\tau)|}\vec{m}(\tau)\cdot\vec{\sigma},
\end{equation}
where $\sinc(x)=\sin(x)/x$ and $|\vec{m}(\tau)|$ is the norm of the vector $\vec{m}(\tau)$. Using expression (\ref{interad}), the 
evolution operator in the Schr\"{o}dinger picture Eq.~(\ref{evolh}) becomes
\begin{equation}\label{ufin}
U(\tau)=e^{if(\tau)}[u_0(\tau)-i\vec{u}(\tau)\cdot\vec{\sigma}], 
\end{equation}
where $f(\tau)$ is defined in equation (\ref{f}) and $u_0(\tau)$ as well as the components of the time-dependent vector $\vec{u}(\tau)=(u_x(\tau),u_y(\tau),u_z(\tau))$  read
\begin{eqnarray}\label{uvector}
\nonumber u_0(\tau)&=&\cos{\frac{\omega_0\tau}{2}}\cos{|\vec{m}(\tau)|}-m_z(\tau)\sin{\frac{\omega_0\tau}{2}}\sinc{|\vec{m}(\tau)|}\\\nonumber
u_x(\tau)&=&\sinc{|\vec{m}(\tau)|}\big[m_x(\tau)\cos{\frac{\omega_0\tau}{2}}-m_y(\tau)\sin{\frac{\omega_0\tau}{2}}\big]\\\nonumber
u_y(\tau)&=&\sinc{|\vec{m}(\tau)|}\big[m_y(\tau)\cos{\frac{\omega_0\tau}{2}}+m_x(\tau)\sin{\frac{\omega_0\tau}{2}}\big]\\\nonumber
u_z(\tau)&=&m_z(\tau)\sinc{|\vec{m}(\tau)|}\cos{\frac{\omega_0\tau}{2}}+\sin{\frac{\omega_0\tau}{2}}\cos{|\vec{m}(\tau)|}.\\
&&
\end{eqnarray}From the unitarity of (\ref{ufin}), the quasienergies $\varepsilon_{\pm}(\kappa)$ are then given by
\begin{equation}\label{quasienergies}
\varepsilon_{\pm}(\kappa)=\pm\frac{1}{2\pi}\tan^{-1}\Bigg(\frac{\sqrt{1-u^2_0(\kappa,2\pi)}}{u_0(\kappa,2\pi)}\Bigg),
\end{equation} where we have again made explicit the momentum dependence. 
Therefore, the calculation of the quasienergy spectrum reduces to finding $u_0(2\pi)$, which in turn amounts to calculate both $|\vec{m}(2\pi)|$ and $m_z(2\pi)$. 
However it is important to point out that for studying the dynamical behavior of any physical quantity of interest all the equations (\ref{uvector}) need to solved. 
In the following we use the notation $\varepsilon_j$ to denote the approximate quasienergy spectrum obtained by means of a $j$-order truncation of the ME. 
To first-order, the components of the vector $\vec{m}(\tau)$ are found by integration of equation (\ref{interactionbeta}) which leads to
\begin{eqnarray}
\nonumber m_x(\tau)&=&\frac{\Lambda\sin\gamma}{2}\Big(\frac{\sin\omega_+\tau}{\omega_+}+\frac{\sin\omega_-\tau}{\omega_-}\Big)\\\nonumber
m_y(\tau)&=&\frac{\Lambda\sin\gamma}{2}\Big(\frac{1-\cos\omega_+\tau}{\omega_+}+\frac{1-\cos\omega_-\tau}{\omega_-}\Big)\\\nonumber
m_z(\tau)&=&-\Lambda\cos\gamma\sin\tau.\\
\end{eqnarray}
\begin{figure}[ht]
\begin{center}
\includegraphics[height=6.5cm]{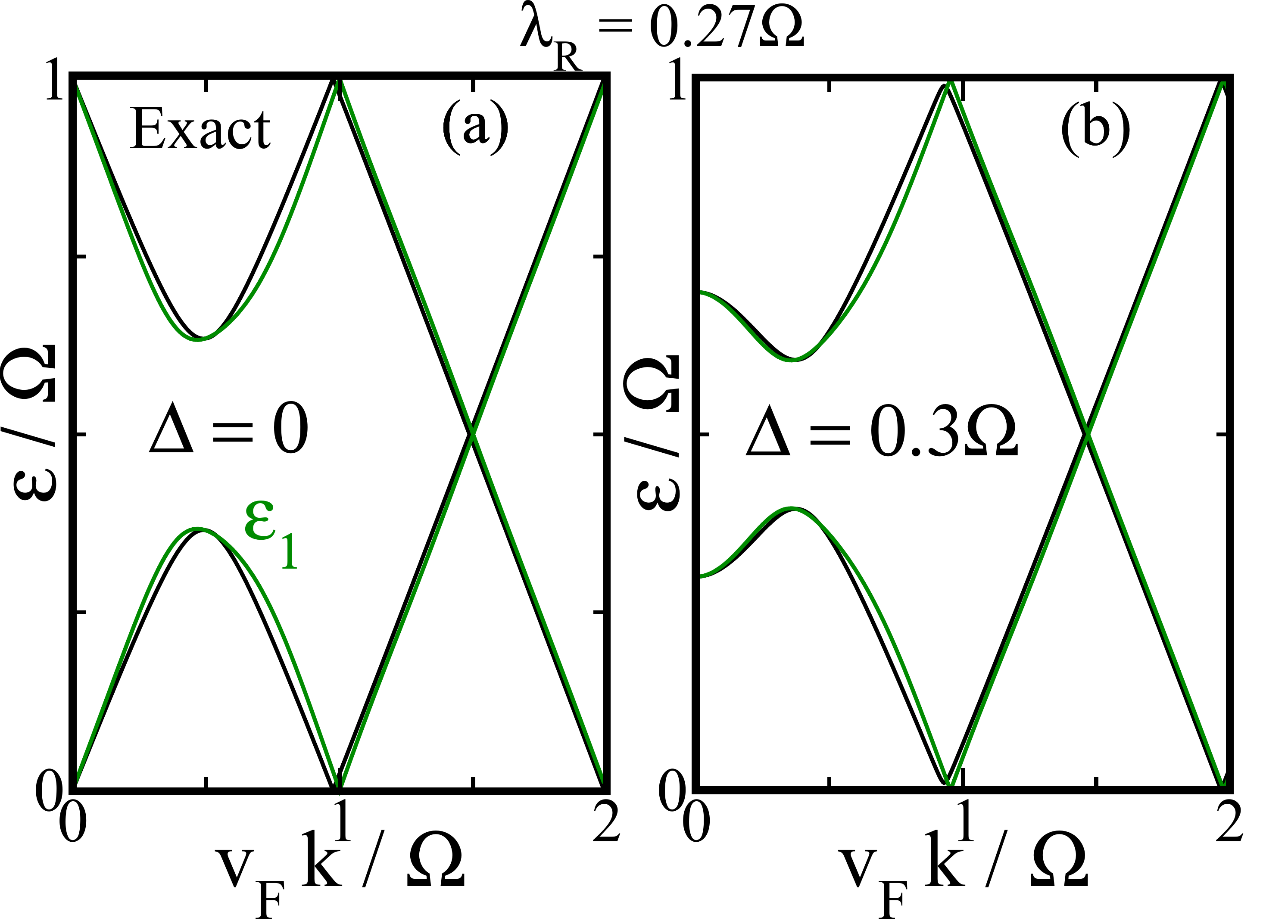}
\caption{\label{fig:figure1}(Color online) The quasienergy spectra obtained by means of an exact numerical diagonalization (black line)  and the first-oder ME approximation (green line) 
evaluated at $\lambda_R=0.27\Omega$, well below the boundary value derived from the convergence condition reported in reference. \cite{casas} 
Fixing $\Omega$ as energy scale one finds good qualitative and quantitative agreement between the first-order ME  and the numerical exact solutions within one period as shown in 
panel a (b) for vanishing (finite) intrinsic spin-orbit contribution $\Delta$.}
\end{center}
\end{figure}In order to simplify the notation we have introduced the shifted frequencies $\omega_\pm=\omega_0\pm 1$.
Notice that for $\tau=2\pi$, the component $m_z(\tau)$ vanishes. Therefore, the modulus $m(2\pi)\equiv m$ becomes
\begin{eqnarray}\label{conv}
m&=&\frac{4\pi\Lambda\kappa}{\omega_+}|\sinc(\pi\omega_-)|.
\end{eqnarray}
and we get to first-order in the interaction strength 
\begin{equation}
\varepsilon_{\pm}(\kappa)=\pm\frac{1}{2\pi}\tan^{-1}\Bigg(\frac{\sqrt{1-\cos^2\pi\omega_0\cos^2 m}}{\cos\pi\omega_0\cos m}\Bigg),
\end{equation}
whereas taking into account the second-order contributions leads to rather lengthier expressions since now we have $|\vec{m}(2\pi)|=\sqrt{m^2_z+m^2_{||}}$, with
\begin{eqnarray}
\nonumber|m_{||}|&=&4\Lambda\kappa\Bigg|\frac{\pi\sinc(\pi\omega_-)}{\omega_+}-\frac{2\Lambda\delta\sinc{(\pi\omega_0)}}{\omega_0(\omega^2_0-4)}\Bigg|\\
m_{z}&=&\frac{4\Lambda^2\kappa^2}{\omega_+\omega_-}\Bigg[\frac{\pi}{\omega_0}-\frac{\sin(2\pi\omega_0)}{\omega_+\omega_-}\Bigg].\\\nonumber
&&
\end{eqnarray}

In the appendix we summarize the results for the contributions to the ME up to the third-order.
\begin{figure}[ht]
\begin{center}
\includegraphics[height=6.5cm]{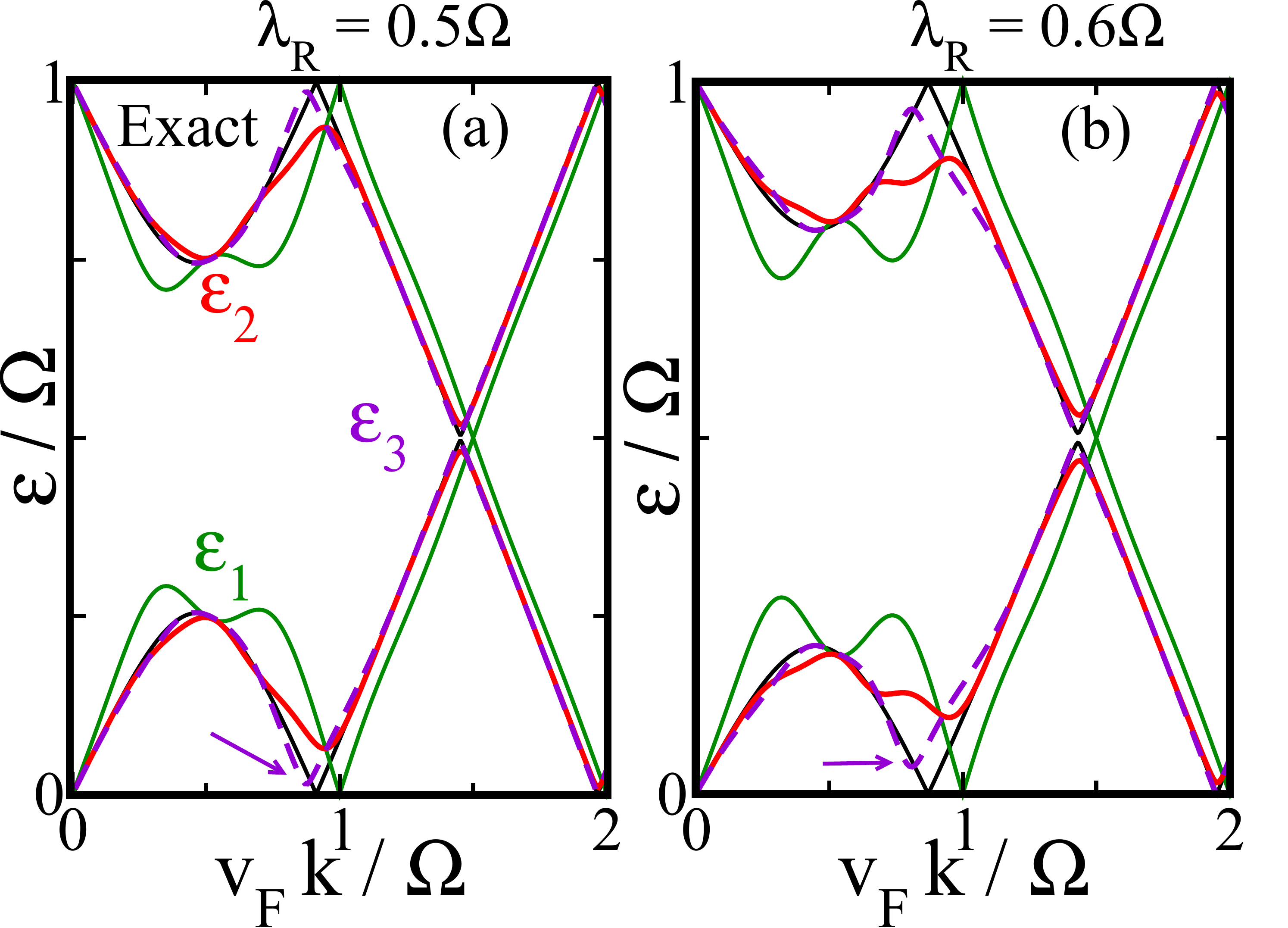}
\caption{\label{fig:figure3} (Color online) Quasi-energy spectra for exact (black, continuos line), first (green, continous, line), second (red, continous line) and third 
(violet, dashed line) ME  solutions for vanishing $\delta=0$ mass term. In this strong coupling regime with values of the effective coupling amplitude below the one given by the 
convergence criteria $\lambda_R=0.78\Omega$ reported in\cite{moan,casas} the third-order ME fits very well the exact result (see discussion in the main text).}
\end{center}
\end{figure} 
In figure~\ref{fig:figure1} we show the first-order ME quasienergy spectrum corresponding to $\lambda_R=0.27\Omega$ (i.e., well below the critical value $\lambda_R=0.78\Omega$). 
Panel a (b) shows the result for zero (finite) intrinsic spin-orbit contribution. In both panels of figure \ref{fig:figure1} the green line corresponds to the 
first-order ME whereas the black line is the result from the exact numerical diagonalization. Thus, for this small coupling we see that even the 
first-order ME solution fits qualitatively and quantitatively well the results from the exact numerical solution and describes properly the gap opening at finite momenta 
reported in references \cite{oka,wu,foa}, with the largest gap located at $\omega_0=1/2$. 
To approach the convergence bound $\lambda_R=0.78\Omega$, higher order contributions to the ME need to be 
included. This is shown in figure \ref{fig:figure3} where we present the exact numerical solution (black continous) along with the first (green, continous line), second (red, continous line) and third 
(violet, dashed line) order one-period ME approximation for $\lambda_R=0.5\Omega$ (a) and $\lambda_R= 0.6\Omega$ (b). As the convergence is not changed by $\Delta$ we restrict ourselves, 
without loss of generality, to the case $\Delta=0$. From panel (a) in \ref{fig:figure3} we find that up to $\lambda_R=0.5\Omega$ only third-order ME fits well the numerically exact solution, with minor 
deviations in the vicinity of the zero of the exact solution. This zero of the quasienergies appears in the vicinity of  $v_F k=\Omega$. As it is depicted 
by the arrows, this discrepancy is more noticeable in panel (b). 
\begin{figure}[ht]
\begin{center}
\includegraphics[height=6.5cm]{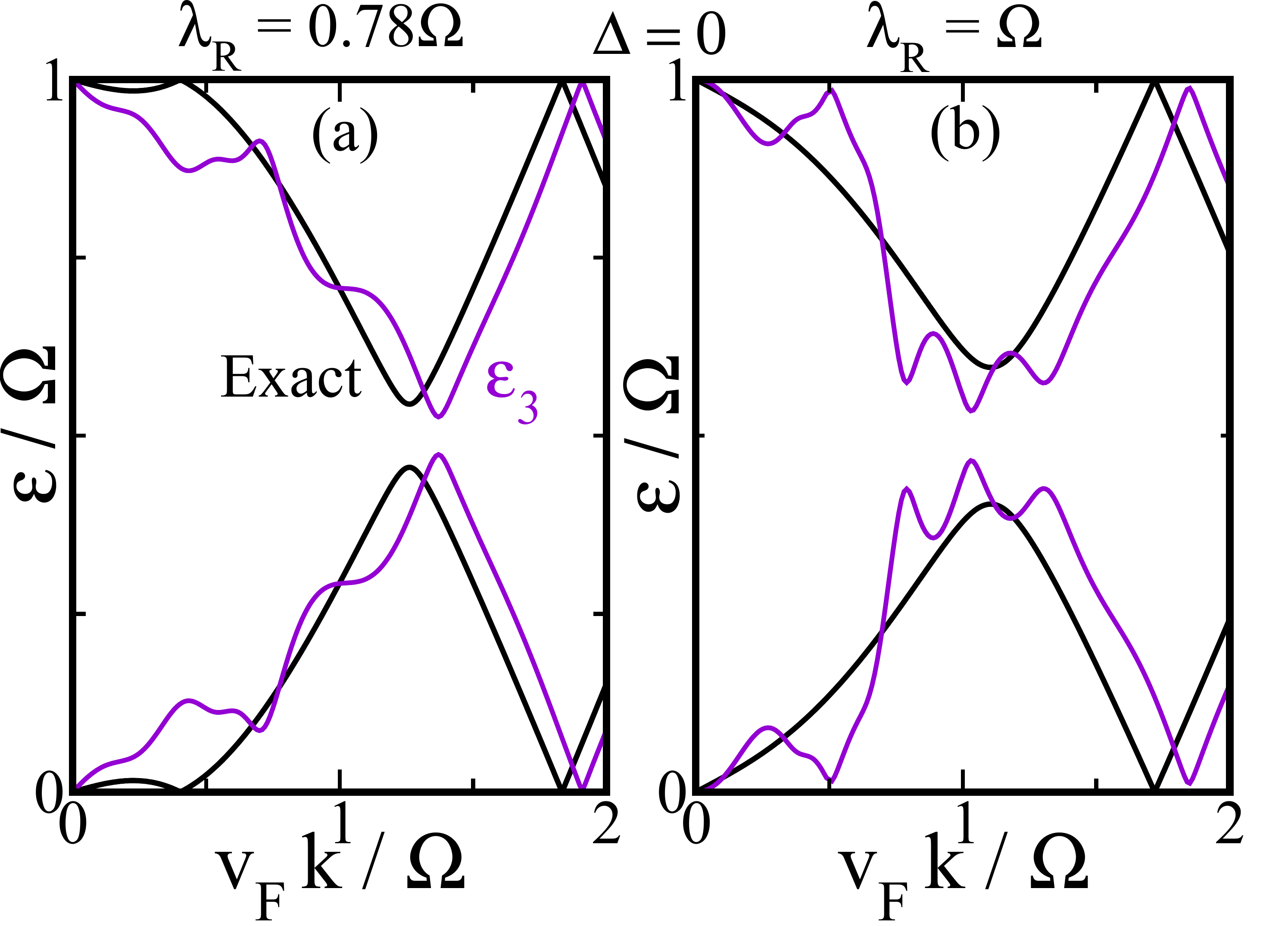}
\caption{\label{fig:figure4} (Color online) Quasi-energy spectra for exact (black line) and third-order (violet line) 
ME  solutions, for vanishing ($\delta=0$) mass term. We notice that at $\lambda_R=0.78\Omega$ the third-order ME considerably deviates from the exact result
as more terms in the perturbative expansion are required to fit the exact solution. However, at larger $\lambda_R=\Omega$(panel b), the ME does not converge and addition of more terms 
would not improve the approximated quasienergy spectrum.}
\end{center}
\end{figure}

On the other hand,  in panel (a) of  figure \ref{fig:figure4} we explore the RSO coupling regime at the boundary $\lambda_R=0.78\Omega$ whereas in panel (b) we set $\lambda_R=\Omega$, i.e., beyond the 
convergence domain. In the left panel we notice that even at the convergence boundary the truncated ME solution  $\varepsilon_3$ (violet, continuous line) does not properly describe the locations of the gap openings from 
the exact solution (black, continuos line), and also shows a wavelike behavior indicating that more terms are needed to get a better approximation. This is a consequence of the perturbative nature of the approach. Still, 
we find that the ME solution qualitatively ``follows'' the exact solution. For coupling values beyond the convergence bound (right panel) the discrepancies are more obvious because even the full ME 
solution breaks down. In contradistinction to the improvements resulting from adding of more terms to the ME, due to the convergence restriction, the coupling regime 
$\lambda_R>0.78\Omega$ is not suitable to be treated by means of the ME approach.

Although the convergence restriction restrains the applicability of the ME it is important to remark that one of its main advantages over the numerical Fourier-mode counterpart is 
that we can get valuable analytical information on the momentum dependence of the gaps by just
finding the maxima of the first-period quasienergy $\varepsilon_+\equiv\varepsilon(\kappa)$. For instance, within the low coupling limit the first-order ME expressions given in equation 
Eq.~(\ref{quasienergies}) appropriately gives the expected result for the gaps at $2\kappa=1$. Yet, we 
notice in  panel (a) of figure \ref{fig:figure4} that for large values of the coupling constant, the gap is noticeable shifted from $2\kappa=1$. Its behavior is
 obtained from the maxima of
\begin{equation}\label{quasienergies11}
\varepsilon(\kappa)=\frac{1}{2\pi}\tan^{-1}\Bigg(\frac{\sqrt{1-u^2_0(\kappa)}}{u_0(\kappa)}\Bigg),
\end{equation}
that leads to the following expressions
\begin{eqnarray}
&&\nonumber\Bigg(\frac{2u_0^2-1}{\sqrt{1-u^2_0}}\Bigg)\frac{\partial u_0}{\partial\kappa}=0,\\\nonumber
&&\frac{1}{2\pi}\Bigg[\frac{u_0[1+2(1-u_0^2)]}{(1-u^2_0)^{3/2}}\Bigg(\frac{\partial u_0}{\partial\kappa}\Bigg)^2+\frac{2u_0^2-1}{\sqrt{1-u^2_0}}\frac{\partial^2 u_0}{\partial\kappa^2}\Bigg]<0,\\\nonumber
&&
\end{eqnarray}
where, to simplify  the notation, we have set $u_0(\kappa,2\pi)=u_0$.
The first relation implies either $u_0^2=1/2$ and/or $\frac{\partial u_0}{\partial\kappa}=0$. However, substitution of $u_0^2=1/2$ on the second condition leads to the contradiction 
$(2\frac{\partial u_0}{\partial\kappa})^2<0$. Then, the maxima are found from the (in principle) simplified conditions $\frac{\partial u_0}{\partial\kappa}=0$, $\frac{\partial^2 u_0}{\partial\kappa^2}<0$.

\section{Discussion}
The fact that changing to the interaction representation  improves the
convergence of the ME is certainly not a peculiarity of the model considered here. 
For example, Pechukas and Light\cite{pechukas} showed 
that this is also true for the paradigmatic example of a  driven quantum harmonic 
oscillator. In particular, for the linearly driven quantum harmonic oscillator the 
ME terminates at second-order, and thus no truncation is necessary. Hence, only convergence issues have to be dealt with. 
In addition, for a two-level driven problem the same authors show that the 
$m$-th order ME solution gives a more accurate description of the dynamics 
than perturbation theory to the same order of approximation.
This higher degree of accuracy is associated to the unitarity preserving property of the ME approach. Salzman\cite{salzman} and 
Fernandez\cite{fernandez2} also deal with the driven quantum  harmonic 
oscillator and conclude that the ME solution will always converge for times within the first period 
$0 < t < T$. This problem admits an exact analytical solution and the authors 
show that the ME solution gives a proper account of the dynamics apart for 
values of the driving frequency near resonances. 
Our results indicate that, within its convergence domain, the truncated ME provides a suitable description of the dynamical behavior of physical quantities within the time window $0 < t < T$. However, 
due to the momentum independence of the restriction $\lambda_R=0.78\Omega$, we found that intermediate values of the coupling ($0.5\Omega\leq\lambda_R\leq0.6\Omega$), the quasienergy spectrum is properly described even at resonances by the third-order 
truncated ME. Concerning energy scales the 
value of the intrinsic and extrinsic spin-orbit coupling parameters $\Delta$ and $\lambda_R$ in graphene have been obtained  by tight binding \cite{macdonald} and  first principle 
calculations \cite{yao,mitra}. They gave estimates in the range $10^{-6}-10^{-5}~{\rm eV}$, much smaller than any other energy scale  in the problem 
(kinetic, interaction and disorder). However, the RSOC strength has recently been reported\cite{rashbaexp} from the band splitting to be of order $2\lambda_R\approx225~{\rm meV}$. 
Using this value and the boundary (in full dimensional form) $2\lambda_R=\pi\hbar\Omega/2$ and putting $\Omega=2\pi\nu$ we get the 
corresponding frequency to be $\nu\approx35~ {\rm THz}$.
In addition, we have seen that comparing the 
semi-analytical to the exact numerical approach a key feature of 
the dynamical description by means of the truncated ME approach is that 
explicit formulae are obtained for the quasienergy spectrum. 
Therefore, the analysis of the gap openings in the low and intermediate 
coupling regime can be given a semi-analytical treatment. This feature might provide useful information on the underlying 
physical processes leading to a better understanding of the nature of 
these important features of the quasienergy spectrum in more complicated 
setups. 
\section{Conclusions}
\noindent In this work we have  described semi-analytically the quasi-energy spectrum of 
charge carriers in graphene under ac-driven RSOC interaction by means of the 
ME approach. We have shown that within the Schr\"odinger picture the 
ME is only applicable within a small neighborhood of the Dirac point and makes it non suitable to describe the induced gap openings at finite momenta. This difficulty is overcome by
changing to the interaction picture where the convergence domain becomes momentum independent and just restricts the effective coupling to values $\lambda_R\leq0.78\Omega$. 
Although using this formulation the truncated ME evolution operator violates the stroboscopic property, we found that for low values of the coupling constant the truncated first-order ME solution fits 
very well the exact quasienergy spectrum. In addition, for values of the coupling constant up to $\lambda_R=0.5\Omega$ the spectrum is properly described by the third-order ME solution. Evaluation at the boundary 
$\lambda_R=0.78\Omega$ shows the need to include more terms, whereas for values such as $\lambda_R=\Omega$ the Magnus approximation breaks down. 
As shown in previous works\cite{fernandez,salzman} the ME always converges in $0 < t < T$, and therefore our results for the one-period quasienergy spectrum are in agreement with these previous 
reports. In contrast to its numerical counterparts, one of the key features of the ME approach is that its analytical nature allows the determination of the momentum dependence of the gaps from which valuable information 
on the position of the resonant processes can be obtained.  In summary, we 
have shown that implementation of the ME approach provides a valuable tool to get physical insight into the mean features of a driven problem and it would shed light on the 
description of more complicated driven problems, provided good care is paid on determining the appropriate parameter ranges as we have discussed.

{\bf Acknowledgments.}
We thank M.~W. Wu and Y. Zhou for useful correspondence. A. L. Acknowledges fruitful discussions with B. Santos.
This work has been supported by Deutsche 
Forschungsgemeinschaft via GRK 1570.

\section*{Appendix}
\subsection*{Derivation of second and third-order contributions to the ME in the interaction picture}
The adimensional dynamical generator in the interaction picture is written in terms of the Pauli matrices as
\begin{equation}
V_I(\tau)=\Lambda\cos\tau\big[\sin\gamma\big(\cos\omega _0 \tau \sigma_x-\sin\omega _0 \tau\sigma_y\big)-\cos\gamma\sigma_z\big].
\end{equation}
Using the symplifying notation $V_I(\tau_j)=V_j$, and denoting the second and third-order contribution to the ME respectively as $M_2$ and $M_3$,
one has
\begin{eqnarray}\label{a1}
\nonumber M_2(\tau)&=&\frac{-i^2}{2}\int_0^\tau d\tau_1\int^{\tau_1}_0\big[V_1,V_2\big] d\tau_2\\\nonumber
M_3(\tau)&=&\frac{-i^3}{6}\int_0^\tau d\tau_1\int_0^{\tau_1}d\tau_2\int^{\tau_2}_0\Big(\big[V_1,\big[V_2,V_3\big]\big]\\\nonumber
&&+\big[V_3,\big[V_2,V_1\big]\big]\Big) d\tau_3.\\
\end{eqnarray}
Performing the integrations, and evaluating at one period, i.e. $\tau=2\pi$,  we obtain for the second-order ME term the result
\begin{eqnarray}\label{teste}
\nonumber M_2(2\pi)&=&\frac{\Lambda ^2}{2 \big(\omega _0^2-4\big)\big(\omega _0^2-1)^2}\Bigg\{4 \sin2\gamma \sin\pi\omega_0 \\\nonumber
&&\times\big(\sigma_x\cos\pi\omega_0 -\sigma_y \sin\pi\omega_0\big)\big(\omega _0^2-1\big)^2\\\nonumber
&&-\omega _0\sigma_z \Big[\pi  \big(\omega _0^2-1\big)-\omega _0\sin2\pi\omega_0\Big]\\\nonumber
&&\times\big(\omega _0^2-4\big)\big(1-\cos2\gamma\big)\Bigg\}.\\
\end{eqnarray}
Therefore, to second order we get now a finite $\sigma_z$ component that shifts the zeroes of the quasienergy spectrum as shown in figures~\ref{fig:figure3} and \ref{fig:figure4}. From this expression we 
explicitly see the appeareance 
of another resonance at $\Omega = 2 \Omega_0$, in addition to that corresponding to $\Omega=\Omega_0$, which is properly described by the first-order ME solution as it was explicitly worked out in section IV. This additional 
resonant contribution weights the higher momentum gap openings that are shown in figures \ref{fig:figure1}-\ref{fig:figure3}, yet they are small in amplitude as compared to the one-photon resonance leading to the largest gap 
opening. This is in agreement with the Fourier-mode expansion solution for moderate values of the effective coupling strength $\Lambda$.

In order to use the ME solution for values close to the convergence boundary given by {\it Casas},\cite{casas} we have also evaluated the first-period third-order contribution to the ME.  After 
integration up to the first period one finds in this case the results
\begin{small}
\begin{eqnarray}\label{test}
\nonumber M_3(2\pi)&=&\frac{\Lambda ^3 \omega _0 \sin\gamma}{6\big(\omega _0^2-1\big)^3 \big(\omega _0^2-4\big)\big(\omega _0^2-9\big)}\\\nonumber 
&&\times\Bigg\{12 \sin2\gamma \sin2\pi\omega_0 \big(\omega _0^2-9\big) \big(\omega _0^2-1\big)^2\sigma_z \\\nonumber
&&+\big(\sigma_x\cos\pi\omega_0 -\sigma_y \sin\pi\omega_0\big)\big(\omega _0^2-4\big)\\\nonumber
&&\times\Big[\omega _0\cos2\gamma\Big(3\omega_0  \sin\pi\omega_0 \big(2+11\big(3- \omega _0^2\big)\big)\\\nonumber
&&+\big(\omega _0^2-9\big)\Big(6 \pi \cos\pi\omega_0 \big(\omega _0^2-1\big)-\omega_0\sin3\pi\omega_0\Big) \Big)\\\nonumber
&&-\big(\omega _0^2-9\big)\Big(6 \pi \cos\pi\omega_0 \big(\omega _0^2-1\big)-\omega^2_0\sin3\pi\omega_0\Big) \Big)\\\nonumber
&&-3 \sin\pi\omega_0 \big(16+3 \omega _0^2+5 \omega _0^4\big)\Big]\Bigg\},\\
\end{eqnarray}
\end{small}which, as expected, contains higher order photon resonant contributions. Continuing this perturbative treatment, larger domains of values the coupling strength are reacheable but the  addition of 
higher order terms lead to rather lengthier expressions and go beyond the main focus of the present work.

\end{document}